\documentclass[conference]{IEEEtran}
\IEEEoverridecommandlockouts

\usepackage{cite}
\usepackage{amsmath,amssymb,amsfonts}
\usepackage{graphicx}
\usepackage{optidef}
\usepackage{textcomp}
\usepackage{xcolor}
\usepackage{multirow}
\usepackage{algorithm}
\usepackage{algpseudocode}
\usepackage{float}
\usepackage{booktabs} 
\usepackage{setspace}

\usepackage{siunitx} 
\usepackage{subfig}
\usepackage[draft=true]{hyperref}
\hypersetup{bookmarks=false}
\usepackage[left=0.625in,right=0.625in,top=0.75in,bottom=1.05in]{geometry}
\graphicspath{ {./figures/} }
\def\BibTeX{{\rm B\kern-.05em{\sc i\kern-.025em b}\kern-.08em
    T\kern-.1667em\lower.7ex\hbox{E}\kern-.125emX}}
\begin{document}

\title{Partially-Observable Transmission Control for UAV-Enabled Federated Learning in IoT Networks}

\author{
\IEEEauthorblockN{Masoud Ghazikor, Zhou Ni, Morteza Hashemi}
        \IEEEauthorblockA{Department of Electrical Engineering and Computer Science, University of Kansas,
        }
}

\maketitle

\begin{abstract}
Uncrewed aerial vehicle (UAV)-enabled federated learning (FL) can provide flexible, on-demand edge intelligence for large-scale IoT deployments, but operating in shared unlicensed bands makes uplink update delivery interference-coupled and unreliable. In this paper, we develop a packet-level transmission framework that captures buffer overflow, delay violations, and transmission errors, and uses the resulting packet delivery ratio (PDR) to represent partial-update reception through a packetized, Bernoulli-masked FL aggregation process. We then formulate a fairness-consensus bilevel (FCB) optimization that jointly controls (i) transmission thresholds to maximize the average PDR while reaching consensus under partial observability and (ii) transmission powers to improve the worst PDR and enforce fairness across IoT learners. To solve this problem, we propose an alternating FCB optimizer composed of a consensus-based threshold controller (CTC), which drives the IoT learners toward a PDR-efficient consensus on transmission thresholds, and a fairness-based power controller (FPC), which updates transmission powers to improve the worst PDR and ensure fairness under the resulting consensus thresholds. Numerical results on CNN-based FL tasks show that the FCB optimizer improves FL aggregation and training performance by enhancing packet-level update delivery, consistently outperforming baseline transmission policies.
\end{abstract}


\renewcommand\thefootnote{}\footnotetext{The material is based upon work supported by NSF grants 1955561, 2212565, 2323189, 2434113, and 2514415.}

\section{Introduction}
Federated learning (FL) has recently emerged as a promising distributed learning framework in wireless networks \cite{ni2026personalized}, where edge devices collaboratively train a shared global model without transmitting raw data. In each communication round, selected devices perform local training and upload their model updates to a central server for aggregation \cite{Ghiasvand-2025-Robust}. This paradigm is particularly attractive for large-scale IoT applications, where data are naturally decentralized across many resource-limited devices and may be sensitive to share. In such settings, an uncrewed aerial vehicle (UAV) can serve as a mobile edge server to provide flexible, on-demand coverage and aggregate updates from geographically dispersed devices \cite{Liaq-2025-Delay}.

In unlicensed wireless environments, transmitting FL updates is inherently unreliable due to buffer overflow, delay constraints, and transmission errors. As a result, updates from IoT learners to the edge server may be partially delivered or lost, which changes the effective aggregation behavior and tightly couples network reliability with learning performance \cite{ni2026pfedwn, Channel_Ghazikor_2024}. However, most wireless FL studies assume that once a learner is scheduled, its update is received reliably, and packet-level losses are either ignored or simplified as idealized learner dropout \cite{mcmahan2017communication, karimireddy2020scaffold, li2020federated, ni2026pfedwn}. For example, FedAvg \cite{mcmahan2017communication} assumes perfect aggregation with error-free model-update delivery. Similarly, communication-efficient methods such as SCAFFOLD \cite{karimireddy2020scaffold} and FedProx \cite{li2020federated} primarily address heterogeneity in client data and optimization while treating the communication channel as reliable. Additionally, recent works study wireless-aware FL by optimizing communication and resource allocation under dynamic bandwidth, fading, and scheduling constraints \cite{zhuansun2024communication,wang2025small,yu2024graph}. Nevertheless, these works mainly improve scheduling, compression, and power allocation while still assuming that scheduled updates are received completely and error-free, rather than explicitly modeling packet erasures or partial-update reception. Moreover, recent studies investigate the over-the-air (OTA) \cite{wang2025adaptive} or UAV-assisted \cite{Fu-2024-Federated} FL to improve communication efficiency in wireless edge networks. However, these works typically focus on joint communication–learning optimization and assume reliable aggregation of transmitted updates, without explicitly accounting for packet erasures or partial-update reception caused by unreliable wireless links.
\begin{figure}[t]
\includegraphics[width=0.9\linewidth]{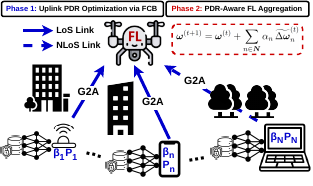}
\centering
\caption{\small UAV-enabled FL operating in unlicensed bands, where multiple ground IoT learners transmit local updates to a UAV server.}
\label{system_model}
\end{figure}

Although there has been research on wireless-aware FL systems, there remains a lack of thorough investigation into \emph{packet-level update delivery to a UAV server}, particularly partial-update reception caused by buffer overflow, delay violations, and interference-coupled uplinks over unlicensed ISM bands, and how these impairments should be incorporated into transmission control across IoT learners. To this end, we propose a comprehensive framework for partially-observable, packet-level transmission control in UAV-enabled FL in ISM bands. The framework models update delivery impairments, including buffer overflow, delay violations, and transmission errors, so that the UAV server may receive partial model updates from IoT learners. Leveraging this model, we propose a fairness-consensus bilevel (FCB, Alg. \ref{alg:FCB}) optimizer to jointly adjust the transmission threshold and transmission power to improve packet delivery ratio (PDR) across IoT learners. The main contributions of this paper are summarized as follows:
\begin{itemize}
    \item We develop a packet-level framework for UAV-enabled FL in ISM bands, capturing buffer overflow, delay violations, and transmission errors to characterize partial model-update reception at the UAV server. Then, we incorporate the PDR into the FL aggregation by modeling packetized updates via a Bernoulli masking process. 
    \item We propose the FCB optimizer to solve the fairness-consensus bilevel optimization problem. In this algorithm, the consensus-based threshold controller (CTC, Alg. \ref{alg:CTC}) addresses the inner problem by adjusting the transmission thresholds to maximize the average PDR across partially-observable learners, ensuring consensus, while the fairness-based power controller (FPC, Alg. \ref{alg:FPC}) addresses the outer problem by tuning the transmission powers to maximize the worst PDR, promoting fairness. 
    \item We demonstrate the effectiveness of the proposed FCB optimizer in improving consensus and fairness across learners, and we quantify how partial-update reception impacts FL aggregation and training performance under different transmission policy settings.
\end{itemize}

The rest of this paper is organized as follows. In Section~\ref{systemmodel}, the system model is introduced. Section~\ref{problemformulation} formulates a fairness-consensus bilevel optimization problem and presents the proposed FCB algorithm. In Section \ref{numericalresults}, numerical results are provided, followed by the conclusion in Section \ref{conclusion}. 

\section{System Model} \label{systemmodel}
We consider a UAV-enabled FL system in which multiple IoT learners transmit their locally trained model updates to a UAV server over ISM bands. The IoT learners operate under practical limitations, including finite buffer sizes, packet delay constraints, and interference from simultaneous uplink transmissions. As a result, local model updates may not be delivered reliably to the UAV server, and some updates may be partially received. Moreover, depending on the propagation environment, each learner may communicate with the UAV server through either a line-of-sight (LoS) or a non-line-of-sight (NLoS) link, as shown in Fig. \ref{system_model}. To capture this behavior, we develop an integrated system model that combines packet-level wireless delivery with FL aggregation. Our envisioned system model leverages components from our prior work \cite{Channel_Ghazikor_2024}, while introducing a novel UAV-enabled FL framework that captures realistic wireless transmission limitations.
\subsection{Communication Model}
\noindent
\textbf{Channel Analysis.} We introduce a distance-based LoS probability model that covers all channel types, including the ground-to-air (G2A) channel. Accordingly, $\mathbb{P}_L(d_i)$ is given by \cite{Kim-2019-Impact}:
\begin{equation}
    \mathbb{P}_{L}(d_i) = \Bigl(1-\frac{\sqrt{2\pi}\eta}{d_i^{V}}\left|Q(\frac{z_{i}}{\eta})-Q(\frac{z_{u}}{\eta})\right|\Bigl)^{d_i^{H}\sqrt{\nu\mu}}.
\end{equation}

Here, $\eta$, $\nu$, and $\mu$ are environmental parameters, and $Q(\cdot)$ denotes the $Q$-function. The horizontal and vertical distances between the learner $i$ and the UAV server $u$ are defined as $d_i^{H} = \sqrt{(x_{i}-x_{u})^2 + (y_{i}-y_{u})^2}$, and $d_i^{V} = z_{i}-z_{u}$. Thus, the 3D link distance is determined by $d_i = \sqrt{{d_i^H}^2 + {d_i^V}^2}$ for all $i \in \left\{n,\boldsymbol{m}\right\}$, where $n \in \boldsymbol{N}$ and $\boldsymbol{m} \subset \boldsymbol{N}$ are the source and the set of interferer learners, respectively.

With transmit power $P_n^t$, the received power on sub-channel $f \in \boldsymbol{F}$ is given by $P_n^r = P_n^t |h_n^f|^2$, where $h_n^f$ denotes the complex channel gain. We decompose $h_n^f$ into $h_n^f = \Tilde{h}_n^f \hat{h}_n^f$, where $\Tilde{h}_n^f$ and $\hat{h}_n^f$ are the small-scale fading coefficient and the square root of the large-scale path loss, respectively. Using the single-slope path loss model in~\cite{Channel_Ghazikor_2024}, we write:
\begin{align}
\hat{h}_n^f = \frac{c}{4\pi d_0 f_c} (\frac{d_0}{d_n})^{0.5\alpha(d_n)} \quad \text{if} \enskip d_n \ge d_0, 
\end{align}
where $d_0$ and $f_c$ denote the reference distance and the carrier frequency, respectively. Moreover, the distance-dependent path loss exponent is modeled as $\alpha(d_n) = \alpha_{L}\mathbb{P}_{L}(d_n) + \alpha_{N}(1 - \mathbb{P}_{L}(d_n))$, where $\alpha_{L}$ and $\alpha_{N}$ are the path loss exponents for LoS and NLoS links, respectively \cite{Channel_Ghazikor_2024}. Under the block fading channel model, the small-scale fading coefficient $\Tilde{h}_{n}^f$ is assumed to follow a Nakagami-$m$ distribution. Thus, its probability density function (PDF) is given by \cite{Ghazikor-2026-Performance}:
\begin{align}
\mathrm{f}_{\Tilde{h}_{n}^f}(x) = \frac{2m_n^{m_n} x^{2m_n-1}}{\Gamma(m_n) \Omega_n^{m_n}} \exp{(-\frac{m_n}{\Omega_n}  x^2)}, 
\end{align}
where $\Gamma(\cdot)$ and $\Omega_n$ are the gamma function and average fading power, respectively. Furthermore, $m_n$ denotes the shape parameter, which is approximated as $m_n \approx \frac{(\exp(2.708\mathbb{P}_{L}(d_n)^2) + 1)^2}{2 \exp(2.708\mathbb{P}_{L}(d_n)^2) + 1}$ using the Rician $K$-factor \cite{Channel_Ghazikor_2024, Ghazikor-2026-Performance}. 

An IoT learner transmits its packet to the UAV server over the best sub-channel $f^\star$, selected as $f^\star = {\arg\max}_{f \in \boldsymbol{F}} \Tilde{h}_n^f \hat{h}_n^f$ if the fading coefficient exceeds the transmission threshold $\beta_n$, i.e., $\Tilde{h}_n^{f^\star} \ge \beta_n$; otherwise, the packet is buffered for future transmission \cite{Guan-2016-ToTransmit}. Accordingly, the cumulative distribution function (CDF) of the Nakagami-$m$ fading at $\beta_n$ is given by:
\begin{align}
\mathcal{F}_{\Tilde{h}_{n}^f}(\beta_n) & = \int_{0}^{\beta_n} \mathrm{f}_{\Tilde{h}_{n}^f}(x) \,dx
= \Tilde{\gamma}_{inc}(m_n, \frac{m_n}{\Omega_n}  \beta_n^2).
\end{align}

Here, $\Tilde{\gamma}_{inc}(\cdot, \cdot)$ denotes the regularized lower incomplete gamma function. Let $|\boldsymbol{F}|$ denote the cardinality of $\boldsymbol{F}$, i.e., the number of available sub-channels. Accordingly, the probability that the IoT learner $n$ transmits a packet in a given time slot under Nakagami-$m$ fading can be expressed as:
\begin{align} \label{murice}
\begin{aligned}
\mu_n(\beta_n) = 1-\mathcal{F}_{\Tilde{h}_n^{f^\star}}(\beta_n) = 1- \Tilde{\gamma}_{inc}(m_n, \frac{m_n}{\Omega_n}  \beta_n^2)^{|\boldsymbol{F}|}. 
\end{aligned}
\end{align}

\noindent
\textbf{Queuing Analysis.}
Time-sensitive model update packets from an IoT learner must be delivered to the UAV server before a deadline. Let $T_n$ denote the queuing delay; under poor channel conditions, transmission may be deferred, and packets with $T_n>T_n^{th}$ are discarded. The delay violation (drop) probability under the time threshold $T_n^{th}$ in M/M/1 queue, denoted by $\mathbb{P}_n^{d}(\beta_n)$, is given by \cite{Channel_Ghazikor_2024, Guan-2016-ToTransmit}:
\begin{align}
\mathbb{P}_n^{d} (\beta_n) \triangleq \mathbb{P}(T_n > T_n^{th}) = \exp{\Big((\lambda_n - \frac{\mu_n(\beta_n)} {T^{s}_n}) T_n^{th} \Big)}, 
\end{align}
where $\lambda_n$ denotes the average packet arrival rate modeled as a Poisson distribution and $T^{s}_n$ is the time slot duration. 

In addition, when poor channel conditions prevent transmission, model update packets may accumulate and cause buffer overflow; once the buffer is full, new arrivals are dropped. Using queuing theory and the Markov chain, the buffer overflow probability can be approximated as \cite{Channel_Ghazikor_2024}:
\begin{align*}
\mathbb{P}_n^{o}(\beta_n) & \approx \frac{(1-\rho_n(\beta_n)) \exp{(\Tilde{b}_n(\rho_n(\beta_n)-1))}}{1-\rho_n(\beta_n) \exp{(\Tilde{b}_n(\rho_n(\beta_n)-1))}},
\end{align*}
where $\rho_n(\beta_n) = \frac{\lambda_nT^{s}_n}{\mu_n(\beta_n)}$ and $\Tilde{b}_n = b_n \eta_n$ denote the offered load and normalized buffer capacity, respectively.

\noindent
\textbf{Interference Analysis.}
We characterize the interference impact on the desired link between the IoT learner $n$ and UAV server by evaluating the transmission error probability. Specifically, given a signal-to-interference-plus-noise ratio (SINR) threshold $\gamma^{th}_n$, an error occurs when $\mathrm{SINR}<\gamma^{th}_n$. Let $I_n^f(\boldsymbol{\beta}_{-n}, \boldsymbol{P}_{-n})$ denote the aggregate interference on sub-channel $f$ from the set of interferers $\boldsymbol{m}$, expressed as \cite{Guan-2016-ToTransmit}:
\begin{align} \label{interference}
I_n^f(\boldsymbol{\beta}_{-n}, \boldsymbol{P}_{-n}) = \sum_{m\in \boldsymbol{N} \backslash n} P_m(\hat{h}_{m}^f \Tilde{h}_{m}^f)^2\alpha_m^f(\beta_m), 
\end{align}
where $\boldsymbol{P}_{-n}\triangleq(P_m)_{m\in\boldsymbol{N}\setminus{n}}$ and $\boldsymbol{\beta}_{-n}\triangleq(\beta_m)_{m\in\boldsymbol{N}\setminus{n}}$ collect the interferers' transmission powers and thresholds, respectively. Let $\alpha_m^f(\beta_m)\in\{0,1\}$ indicate whether interferer $m$ transmits on sub-channel $f$. By leveraging a stochastic geometry approach, the aggregate interference $I_n^f(\boldsymbol{\beta}_{-n}, \boldsymbol{P}_{-n})$ is modeled as a log-normal distribution. Thus, the complementary CDF (CCDF) of $I_n^f(\boldsymbol{\beta}_{-n}, \boldsymbol{P}_{-n})$ is given by: 
\begin{equation}
\begin{aligned}
v_n(x) \triangleq 1 - \mathcal{F}_{I_n^f} (x) = 1-\phi(\frac{\ln(x)-\mu(\boldsymbol{\beta}_{-n}, \boldsymbol{P}_{-n})}{\sigma(\boldsymbol{\beta}_{-n}, \boldsymbol{P}_{-n})}).
\end{aligned}
\end{equation}
Here, $\mu(\boldsymbol{\beta}_{-n}, \boldsymbol{P}_{-n})$ and 
$\sigma(\boldsymbol{\beta}_{-n}, \boldsymbol{P}_{-n})$ denotes the location and scale parameters of the log-normal distribution, respectively, as derived in \cite{Channel_Ghazikor_2024}. Furthermore, $\phi(\cdot)$ denotes the CDF of the standard normal distribution. Accordingly, transmission error probability $\mathbb{P}_n^{e}(\boldsymbol{\beta}, \boldsymbol{P})$ is defined as \cite{Guan-2016-ToTransmit}:
\begin{align} \label{perr}
\begin{aligned}
\mathbb{P}_n^{e}(\boldsymbol{\beta}, \boldsymbol{P}) & = \mathbb{P}\Big(\frac{P_n(\hat{h}_n^f \Tilde{h}_n^f)^2}{I_n^f(\boldsymbol{\beta}_{-n}, \boldsymbol{P}_{-n}) + N} < \gamma^{th}_n \Big) \\ &
= \int_{\beta_n}^{\infty} \mathrm{f}_{\Tilde{h}_n^f} (x) v_n\Big(\frac{P_n(\hat{h}_n^f)^2}{\gamma^{th}_n}x^2-N \Big) dx,
\end{aligned}
\end{align}
where $N=kTW$ is the noise power, with $k$, $T$, and $W$ denoting the Boltzmann constant, temperature, and bandwidth.

\noindent
\textbf{PDR.}
According to the queuing and interference analyses, the PDR of learner $n$ to the UAV server is approximated as \cite{Channel_Ghazikor_2024}:
\begin{align} \label{Rn}
R_n(\boldsymbol{\beta}, \boldsymbol{P}) \approx [1 - \mathbb{P}_n^{d}(\beta_n) - \mathbb{P}_n^{o}(\beta_n) - \mathbb{P}_n^{e}(\boldsymbol{\beta}, \boldsymbol{P})]^1_0. 
\end{align}

\subsection{Federated Learning Model}
\noindent
\textbf{FL Setup.} We consider a wireless FL system composed of one parameter UAV server and a set of IoT learners $\boldsymbol{N}$. Each learner $n$ stores a private dataset $\mathcal{D}_{n} = \{(x_{i},y_{i})\}_{i=1}^{D_{n}}$, where $D_{n} \triangleq |\mathcal{D}_{n}|$, and the total sample size is $D = \sum_{n = 1}^{|\boldsymbol{N}|}D_{n}$. The learning objective is to minimize the global empirical risk:
\begin{align}
\min_{\boldsymbol{\omega}\in\mathbb{R}^{d}} F(\boldsymbol{\omega}) = \sum_{n=1}^{|\boldsymbol{N}|} \frac{D_n}{D} F_n(\boldsymbol{\omega}).
\end{align}
Here, the local loss is $F_n(\boldsymbol{\omega})
= \frac{1}{D_n}\sum_{(x,y)\in\mathcal{D}_n} \ell(\boldsymbol{\omega};x,y)$ at learner $n$, where $\ell(\cdot)$ denotes the per-sample training loss.

During FL over a wireless network, the training proceeds over synchronous communication rounds indexed by $t = 0,1,2, \dots$. At the beginning of round $t$, the UAV server broadcasts the current global model $\boldsymbol{\omega}^{(t)}$ to all FL learners in the network. Each learner $n \in \boldsymbol{N}$ initializes $\boldsymbol{\omega}_{n}^{t,0} = \boldsymbol{\omega}^{t}$ and performs $e$ local stochastic gradient descent (SGD) steps. Specifically, for the local learning iteration $e = 0, 1,2,\dots,E-1$:
\begin{align}
\boldsymbol{\omega}_{n}^{(t,e+1)} = \boldsymbol{\omega}_{n}^{(t,e)} - \eta\nabla F_{n}(\boldsymbol{\omega}_{n}^{(t,e)}),
\end{align}
where $\eta$ is the local learning rate. After local training, learner $n$ forms the model update, as:
\begin{align}
\Delta\boldsymbol{\omega}_{n}^{(t)} = \boldsymbol{\omega}_{n}^{(t,E)} -\boldsymbol{\omega}^{(t)}.
\end{align}

In contrast to approaches that upload the full local model, we assume only $\Delta\boldsymbol{\omega}_n^{(t)}$ is transmitted over the uplink channel.

\noindent
\textbf{Packetized Uplink \& Packet-based Loss.} In wireless systems, the update transmission is packetized. We model this by partitioning $\Delta\boldsymbol{\omega}_n^{(t)}\in\mathbb{R}^d$ into $C$ contiguous packets:
\begin{align}
\Delta\boldsymbol{\omega}_n^{(t)}
=\big[\Delta\boldsymbol{\omega}_{n,1}^{(t)},\dots,\Delta\boldsymbol{\omega}_{n,C}^{(t)}\big].
\end{align}

As defined in Eq.~\eqref{Rn}, the PDR denotes the probability that a packet from learner $n$ is successfully received at the UAV server, and is used as the packet-level success probability in FL since each local update is split into $C$ packets.

For each communication round $t$, the reception indicator of packet $c \in \{1,\dots,C\}$ from learner $n$ is modeled as
$b_{n,c}^{(t)} \sim \mathrm{Bernoulli}(R_n(\boldsymbol{\beta},\boldsymbol{P}))$,
where $b_{n,c}^{(t)} = 1$ indicates successful reception at the UAV server and $b_{n,c}^{(t)} = 0$ indicates packet loss. Under the standard assumption of independent packet transmissions, the number of successfully received packets follows a binomial distribution with mean $C R_n(\boldsymbol{\beta},\boldsymbol{P})$. Accordingly, the UAV server receives a partially corrupted local update:
\begin{align}
\widetilde{\Delta\boldsymbol{\omega}}_n^{(t)} =
\big[
b_{n,1}^{(t)}\Delta\boldsymbol{\omega}_{n,1}^{(t)},\dots,
b_{n,C}^{(t)}\Delta\boldsymbol{\omega}_{n,C}^{(t)}
\big].
\end{align}
Equivalently, letting $\mathbf{B}_n^{(t)}$ denote the induced block-diagonal masking operator, we write
$\widetilde{\Delta\boldsymbol{\omega}}_n^{(t)}
=
\mathbf{B}_n^{(t)}\Delta\boldsymbol{\omega}_n^{(t)}$. After collecting uplink transmissions from all learners, the UAV server updates the global model using the received updates. With standard data-size weighting $
\alpha_n=\frac{D_n}{D}$, the global aggregates the model updates from IoT learners based on:
\begin{align}\label{agg}
\boldsymbol{\omega}^{(t+1)}
=
\boldsymbol{\omega}^{(t)}
+
\sum_{n\in\boldsymbol{N}}
\alpha_n\,\widetilde{\Delta\boldsymbol{\omega}}_n^{(t)}.
\end{align}

\section{Problem Formulation \& Proposed Solution} \label{problemformulation}
\subsection{Fairness-Consensus Bilevel (FCB) Optimization Problem}
In our previous work~\cite{Channel_Ghazikor_2024}, we showed that maximizing only the average PDR can lead to an unbalanced operating point in which a few learners achieve high rates while others become bottlenecks. Moreover, in the considered FL system, a higher PDR improves the reliability of model update delivery to the UAV server, thereby reducing aggregation loss under partial reception and improving training performance. We therefore seek a design that (i) tunes the transmission threshold vector $\boldsymbol{\beta}$ to maximize the average PDR while ensuring consensus, and (ii) adjusts the transmission power vector $\boldsymbol{P}$ to maximize the worst PDR while satisfying fairness. To capture this structure, we formulate the FCB optimization problem in which the \emph{inner} level computes a \emph{consensus} solution for $\boldsymbol{\beta}$ under a fixed $\boldsymbol{P}$, while the \emph{outer} level enforces \emph{fairness} by tuning $\boldsymbol{P}$ based on the inner-level consensus values, as follows:
\begin{subequations}
\begin{align}
\max_{\boldsymbol{P}} & \ \min_{n \in \boldsymbol{N}} \ R_n(\boldsymbol{\beta}^{\star}, \boldsymbol{P}) \label{obj:outer} \\
\text{s.t. } 
& \ P_n^{\min} \le P_n \le P_n^{\max}, \quad \forall n \in \boldsymbol{N}, \label{con:outer-1} \\
& \ J(\boldsymbol{\beta}^{\star}, \boldsymbol{P}, \boldsymbol{w}) \ge \zeta, \label{con:outer-2} \\
& \ \boldsymbol{\beta}^{\star} \in \arg \max_{\boldsymbol{\beta}} \ \frac{1}{|\boldsymbol{N}|} \sum_{n=1}^{|\boldsymbol{N}|} R_n(\boldsymbol{\beta}, \boldsymbol{P}) \label{obj:inner} \\
& \ \text{s.t. } \ 0 < \beta_n \le \beta^{\max}_n, \quad \forall n \in \boldsymbol{N}, \label{con:inner-1} \\
& \qquad \beta_n - \boldsymbol{\beta}^{\star}[n] = 0, \quad \forall n \in \boldsymbol{N}. \label{con:inner-2}
\end{align}
\end{subequations}

In the outer level \eqref{obj:outer}--\eqref{con:outer-2}, we optimize $\boldsymbol{P}$ within the per-learner limits to maximize the minimum PDR. In addition, fairness is enforced through the weighted Jain's index \cite{Sediq-2013-Optimal}:
\begin{equation}
J(\boldsymbol{\beta}, \boldsymbol{P}, \boldsymbol{w}) =
\frac{\left(\sum_{n\in\boldsymbol{N}} w_n \, R_n(\boldsymbol{\beta},\boldsymbol{P})\right)^2}
{\left(\sum_{n\in\boldsymbol{N}} w_n\right)\left(\sum_{n\in\boldsymbol{N}} w_n \, R_n^2(\boldsymbol{\beta},\boldsymbol{P})\right)}.
\label{eq:weighted_jain}
\end{equation}

The outer objective depends on $\boldsymbol{\beta}^{\star}$ from the inner level \eqref{obj:inner}--\eqref{con:inner-2}, which, for a fixed $\boldsymbol{P}$, determines a consensus threshold vector by maximizing the average PDR subject to $0<\beta_n\le \beta_n^{\max}$, where $\beta_n^{\max}$ is derived from the condition $\mathbb{P}_n^{d} (\beta_n) \le 1$. The consensus constraint \eqref{con:inner-2} enforces agreement by requiring each local threshold $\beta_n$ to match its consensus value $\boldsymbol{\beta}^{\star}[n]$ \cite{Boyd-2011-Distributed}, and the resulting $\boldsymbol{\beta}^{\star}$ is used by the outer level to update $\boldsymbol{P}$ and improve the worst PDR.

\begin{algorithm}
\small
\caption{Fairness-Consensus Bilevel (FCB) Optimizer} \label{alg:FCB}
\begin{algorithmic}[1]
\Function{FCB}{$\psi, \boldsymbol{\beta}^{\max}, \Delta, \zeta, \boldsymbol{w}, \boldsymbol{P}^{\min}, \boldsymbol{P}^{\max}$}
    \State $\boldsymbol{P}^{\star} \gets \boldsymbol{P}^{\max}$, $\boldsymbol{\beta}^{\star} \gets \epsilon$
    \For{$i = 1, \dots, I$}
        \State $\boldsymbol{\beta} \gets \boldsymbol{\beta}^{\star}$, $\boldsymbol{P} \gets \boldsymbol{P}^{\star}$
        \State $\boldsymbol{\beta}^{\star} \gets \Call{CTC}{\frac{\psi}{i}, \boldsymbol{\beta}^{\max}, \boldsymbol{\beta}, \boldsymbol{P}^{\star}}$
        \State $\boldsymbol{P}^{\star} \gets \Call{FPC}{\Delta, \zeta, \boldsymbol{w}, \boldsymbol{P}^{\min}, \boldsymbol{P}^{\max}, \boldsymbol{\beta}^{\star}, \boldsymbol{P}}$
        \If{$\|\boldsymbol{\beta} - \boldsymbol{\beta}^{\star}\| \le \epsilon_{\beta}$ \textbf{and} $\|\boldsymbol{P} - \boldsymbol{P}^{\star}\| \le \epsilon_{P}$}
            \State \textbf{break}
        \EndIf
    \EndFor
    \State \Return $\boldsymbol{\beta}^{\star}, \boldsymbol{P}^{\star}$
\EndFunction
\end{algorithmic}
\end{algorithm}

\subsection{Proposed FCB Optimizer}
We solve the FCB problem via alternating optimization with two coordinated loops: (i) an \emph{inner consensus loop} that updates the threshold vector $\boldsymbol{\beta}$ for a fixed $\boldsymbol{P}$, and (ii) an \emph{outer fairness loop} that updates the power vector $\boldsymbol{P}$ using the consensus thresholds $\boldsymbol{\beta}^{\star}$ to improve the worst PDR. Accordingly, Alg.~\ref{alg:FCB} alternates between two sub-algorithms: 1) CTC (Alg.~\ref{alg:CTC}), which produces $\boldsymbol{\beta}^{\star}$ for the current $\boldsymbol{P}$, and 2) FPC (Alg.~\ref{alg:FPC}), which adjusts $\boldsymbol{P}$ to improve the worst PDR under the fixed $\boldsymbol{\beta}^{\star}$. Initialized with aggressive $\boldsymbol{P}^{\star} \gets \boldsymbol{P}^{\max}$ and $\boldsymbol{\beta}^{\star} \gets \epsilon$, iterations continue until $\|\boldsymbol{\beta}-\boldsymbol{\beta}^{\star}\|\le \epsilon_{\beta}$ and $\|\boldsymbol{P}-\boldsymbol{P}^{\star}\|\le \epsilon_{P}$.

\begin{table*}[!t]
    \centering
    \footnotesize
    \caption{FCB performance vs. transmission policy baselines, reported as ($R_n$, $\beta_n$, $P_n$),
    with optimal results in \textbf{bold}.}
    \begin{tabular}{c|ccccc|c|c|c}
        \toprule
        \text{Policy$\backslash$Index} & $n=1$ & $n=5$ & $n=10$ & $n=15$ & $n=20$ & $\Bar{R}$ & $J(\boldsymbol{\beta},\boldsymbol{P},\boldsymbol{1})$ & $(i, j, k)$ \\ 
        \midrule
        \multirow{1}{*}{{$\text{FCB} (\psi=0.1)$}}
            & $\textbf{(0.90, 1.23, 17)}$ & $\textbf{(0.89, 1.25, 18)}$ & $\textbf{(0.90, 1.24, 18)}$ & $\textbf{(0.89, 1.26, 19)}$ & $\textbf{(0.90, 1.22, 16)}$ & $\textbf{0.90}$ & $\textbf{0.99}$ & $\textbf{(3, 10, 44)}$\\
        \midrule
        \multirow{1}{*}{{$\text{FCB} (\psi=0.4)$}}
            & $\textbf{(0.90, 1.23, 17)}$ & $\textbf{(0.89, 1.25, 18)}$ & $\textbf{(0.90, 1.24, 18)}$ & $\textbf{(0.89, 1.26, 19)}$ & $\textbf{(0.90, 1.22, 16)}$ & \textbf{0.90} & $\textbf{0.99}$ & $\textbf{(3, 62, 44)}$ \\
        \midrule
        \multirow{1}{*}{{$\text{Aggressive} $\cite{Guan-2016-ToTransmit}}}
            & $(0.10,0.72,20)$ & $(0.25,0.81,20)$ & $(0.54,0.99,20)$ & $(0.14,0.71,20)$ & $(0.57,1.01,20)$ & 0.39 & 0.80 & $\text{N/A}$ \\
        \midrule
        \multirow{1}{*}{{$\text{Conservative}$\cite{Channel_Ghazikor_2024}}}
            & $(0.54,1.3,10)$ & $(0.79,1.31,10)$ & $(0.25,1.33,10)$ & $(0.62,1.34,10)$ & $(0.74,1.27,10)$ & 0.49 & 0.81 & $\text{N/A}$ \\
        \bottomrule
    \end{tabular}
    \label{tab:fcb_performace}
\end{table*}

\noindent
\textbf{CTC Sub-algorithm.} Alg.~\ref{alg:CTC} drives learners to a consensus, PDR-efficient threshold vector $\boldsymbol{\beta}^{\star}$. In each iteration, learners update $\beta_n$ in parallel using the latest global vector $\boldsymbol{\beta}$ while treating others' thresholds as fixed, following a consensus-based distributed optimization scheme \cite{Channel_Ghazikor_2024,Boyd-2011-Distributed}. Intermittent participation is captured by a drop probability $\psi_i$ since the network is \emph{partially-observable} in the initial stages, leading to missing threshold updates for a subset of learners. Accordingly, if learner $n$ is non-observable, it skips its threshold update, which is flagged by $\boldsymbol{u}[n]=\mathrm{F}$, and cannot contribute to the global threshold vector $\boldsymbol{\beta}$. As FCB progresses and the interference stabilizes, we reduce intermittency by updating the drop probability across outer iterations (e.g., $\psi_i=\psi/i$), increasing participation as $\boldsymbol{\beta}$ and $\boldsymbol{P}$ converge. When $\boldsymbol{u}[n]=\mathrm{T}$, learner $n$ runs a \emph{Local Coordinate Search (LCS)} sub-algorithm \cite{Channel_Ghazikor_2024} to refine $\beta_n$ and improve $R_n(\boldsymbol{\beta},\boldsymbol{P}^{\star})$. LCS performs a one-dimensional accelerated line search along the $n$-th coordinate while holding other thresholds fixed, producing an update $\boldsymbol{\beta}^{\star}[n]$. Aggregating the parallel updates yields $\boldsymbol{\beta}^{\star}$ and CTC iterates until $\|\boldsymbol{\beta}^{\star}-\boldsymbol{\beta}\|<\epsilon$, while preventing premature stopping under intermittency by ensuring no learner is non-observable for two consecutive iterations (via $\hat{\boldsymbol{u}}$ and $\boldsymbol{u}$). The final $\boldsymbol{\beta}^{\star}$ is passed to the FPC sub-algorithm.

\begin{algorithm}
\small
\caption{Consensus-based Threshold Controller (CTC)} \label{alg:CTC}
\begin{algorithmic}[1]
\Function{CTC}{$\psi_i$, $\boldsymbol{\beta}^{\max}$, $\boldsymbol{\beta}$, $\boldsymbol{P}^{\star}$}
    \State $\boldsymbol{\beta}^{\star} \gets \boldsymbol{\beta}$, $\boldsymbol{u} \gets \text{F}$
    \For{$j = 1, \dots, J$}
        \State $\boldsymbol{\beta} \gets \boldsymbol{\beta}^{\star}$, $\hat{\boldsymbol{u}} \gets \boldsymbol{u}$
        \For{\textbf{each $n \in \boldsymbol{N}$}}  \textbf{in parallel}
            \If{\texttt{random()} $\le \psi_i$} 
            \State $\boldsymbol{u}[n] \gets \text{F}$
            \State \textbf{continue}
            \EndIf
            \State $\boldsymbol{u}[n] \gets \text{T}$, $\boldsymbol{\beta}^{\star}[n] \gets \Call{LCS}{\boldsymbol{\beta}, R_n(\boldsymbol{\beta}, \boldsymbol{P}^{\star}), \boldsymbol{\beta}^{\max}[n]}$
        \EndFor
        \If{$\|\boldsymbol{\beta}^{\star} - \boldsymbol{\beta}\| < \epsilon$ \textbf{and} 
        $ \left( \hat{\boldsymbol{u}}[n], \boldsymbol{u}[n] \right) \neq (\text{F}, \text{F})$}
            \State \textbf{break}
        \EndIf
    \EndFor
    \State \Return $\boldsymbol{\beta}^{\star}$
\EndFunction
\end{algorithmic}
\end{algorithm}

\begin{algorithm}
\small
\caption{Fairness-based Power Controller (FPC)}
\label{alg:FPC}
\begin{algorithmic}[1]
\Function{FPC}{$\Delta$, $\zeta$, $\boldsymbol{w}$, $\boldsymbol{P}^{\min}$, $\boldsymbol{P}^{\max}$, $\boldsymbol{\beta}^{\star}$, $\boldsymbol{P}$}
    \For{$k = 1,\dots,K$}
        \If{$J(\boldsymbol{\beta}^{\star}, \boldsymbol{P}, \boldsymbol{w}) < \zeta$}
            \State $\boldsymbol{P} \gets \Call{JIC}{\Delta, \boldsymbol{w}, \boldsymbol{P}^{\min}, \boldsymbol{P}^{\max}, \boldsymbol{\beta}^{\star}, \boldsymbol{P}}$
            \State \textbf{continue}
        \EndIf

        \State $\hat{n} \gets \arg\min_{n} R_n(\boldsymbol{\beta}^{\star}, \boldsymbol{P})$, $R^{\star} \gets R_{\hat{n}}(\boldsymbol{\beta}^{\star}, \boldsymbol{P}), \boldsymbol{P}^{\star} \gets \emptyset$

        \For{\textbf{each $m \in \boldsymbol{N}$}} \textbf{in parallel}
            \State $\boldsymbol{P}_{m} \gets \boldsymbol{P}$
            \If{$m = \hat{n}$}
                \State $\boldsymbol{P}_{m}[m] \gets \min(\boldsymbol{P}^{\max}[m], \boldsymbol{P}[m]+\Delta)$
            \Else
                \State $\boldsymbol{P}_{m}[m] \gets \max(\boldsymbol{P}^{\min}[m], \boldsymbol{P}[m]-\Delta)$
            \EndIf

            \If{$\min_{n\in\boldsymbol{N}} R_n(\boldsymbol{\beta}^{\star}, \boldsymbol{P}_{m}) > R^{\star}$}
                \State $R^{\star} \gets \min_{n\in\boldsymbol{N}} R_n(\boldsymbol{\beta}^{\star}, \boldsymbol{P}_{m})$, $\boldsymbol{P}^{\star} \gets \boldsymbol{P}_{m}$
            \EndIf
        \EndFor

        \If{$\boldsymbol{P}^{\star} = \emptyset$ \textbf{and} $J(\boldsymbol{\beta}^{\star}, \boldsymbol{P}, \boldsymbol{w}) \ge \zeta$}
            \State \textbf{break}
        \EndIf
        \State $\boldsymbol{P} \gets \boldsymbol{P}^{\star}$
    \EndFor
    \State \Return $\boldsymbol{P}^{\star} \gets \boldsymbol{P}$
\EndFunction
\end{algorithmic}
\end{algorithm}

\noindent
\textbf{FPC Sub-algorithm.} Alg.~\ref{alg:FPC} updates the power vector $\boldsymbol{P}$ for the consensus thresholds $\boldsymbol{\beta}^{\star}$ returned by CTC. At each iteration, FPC first checks whether the fairness constraint is satisfied. If $J(\boldsymbol{\beta}^{\star}, \boldsymbol{P}, \boldsymbol{w}) < \zeta$, the algorithm invokes the \emph{Jain's index correction (JIC)} sub-algorithm to restore the target fairness level. Specifically, JIC perturbs each learner's power by $\Delta$ within its feasible range, evaluates the resulting fairness value, and returns the candidate with the highest improvement in $J(\boldsymbol{\beta}^{\star},\boldsymbol{P},\boldsymbol{w})$. When the fairness constraint is satisfied, FPC identifies the bottleneck learner $\hat{n}$ with the minimum PDR and evaluates one-step candidate power updates for all learners in parallel. The candidate update increases the bottleneck's power by $\Delta$, while decreasing each non-bottleneck's power by $\Delta$, subject to the corresponding power bounds. Among all candidates, FPC selects the one that provides the largest improvement in the minimum PDR. The procedure repeats until no candidate provides further improvement while satisfying the required fairness $\zeta$, and then returns the final $\boldsymbol{P}^{\star}$.

\section{Numerical Results} \label{numericalresults}
\noindent
\textbf{Experimental Setup.} To evaluate the FCB framework, we consider a $100 \times 100\text{m}^2$ network in which 20 learners and a UAV server ($z_u=100$m) are distributed according to a Poisson point process (PPP). Each learner communicates with the UAV server, while the remaining learners are treated as interferers. We use the CIFAR-10 dataset \cite{krizhevsky2009learning} and a ResNet18-based CNN under both independent and identically distributed (IID) and non-IID settings. Each learner performs local training following \cite{caldas2018expanding} with a batch size of $10$. Moreover, the ResNet18 architecture contains $11,173,962$ trainable parameters, corresponding to a model size of approximately $42.63$ MB. To transmit local model updates, each update is divided into multiple packets. Specifically, we consider a payload size of $1200$ bytes per packet, which corresponds to $300$ parameters per packet assuming $32$-bit model parameters. These local model update packets are aggregated according to the standard FedAvg, where partially received updates are masked based on the PDR. For the non-IID settings, we use a Dirichlet distribution~\cite{lin2020ensemble} with $\alpha_d = 0.25$ and $\alpha_d = 0.05$ to model mid and strong heterogeneity, respectively. The experiments are implemented with Pytorch on NVIDIA RTX 3090 GPU. The key simulation parameters are listed in Table \ref{tab:sim_parameters}.

\begin{table}[t]
    \centering
    \caption{Key Simulation Parameters}
    \resizebox{\columnwidth}{!}{%
    \label{tab:sim_parameters}
    \begin{tabular}{lc}
        \toprule
        Category & Notation \& Value \\
        \midrule
        LoS Prob. Model & $\eta = 20$, $\nu = 3 \times 10^{-4}$, $\mu = 0.5$\\
        Channel Model & $\alpha_{L}=2$, $\alpha_{N}=3.5$, $d_0=10$m, $|\boldsymbol{F}|=11$\\
        Queueing Model & $T^{th}_n=80$ms, $T_n^s=5$ms, $\lambda_n=100$, $\tilde{b}_n=50$\\
        SINR Model & $\gamma_n^{th}=10$, $W = 20$MHz, $T = 290$K \\
        FCB Algorithm & $\Delta = 1$dBm, $\zeta=0.99$, $w_n=1$ \\
        FL Setting & $\eta = 0.05$, $C = 37,247$ \\
        \bottomrule
    \end{tabular}
    }
    \label{parameters}
\end{table}

\begin{figure*}[t]
\centering
 \begin{minipage}{0.325\textwidth} 
\subfloat[Test accuracy for CIFR-10 IID case.]{\includegraphics[width=\textwidth]{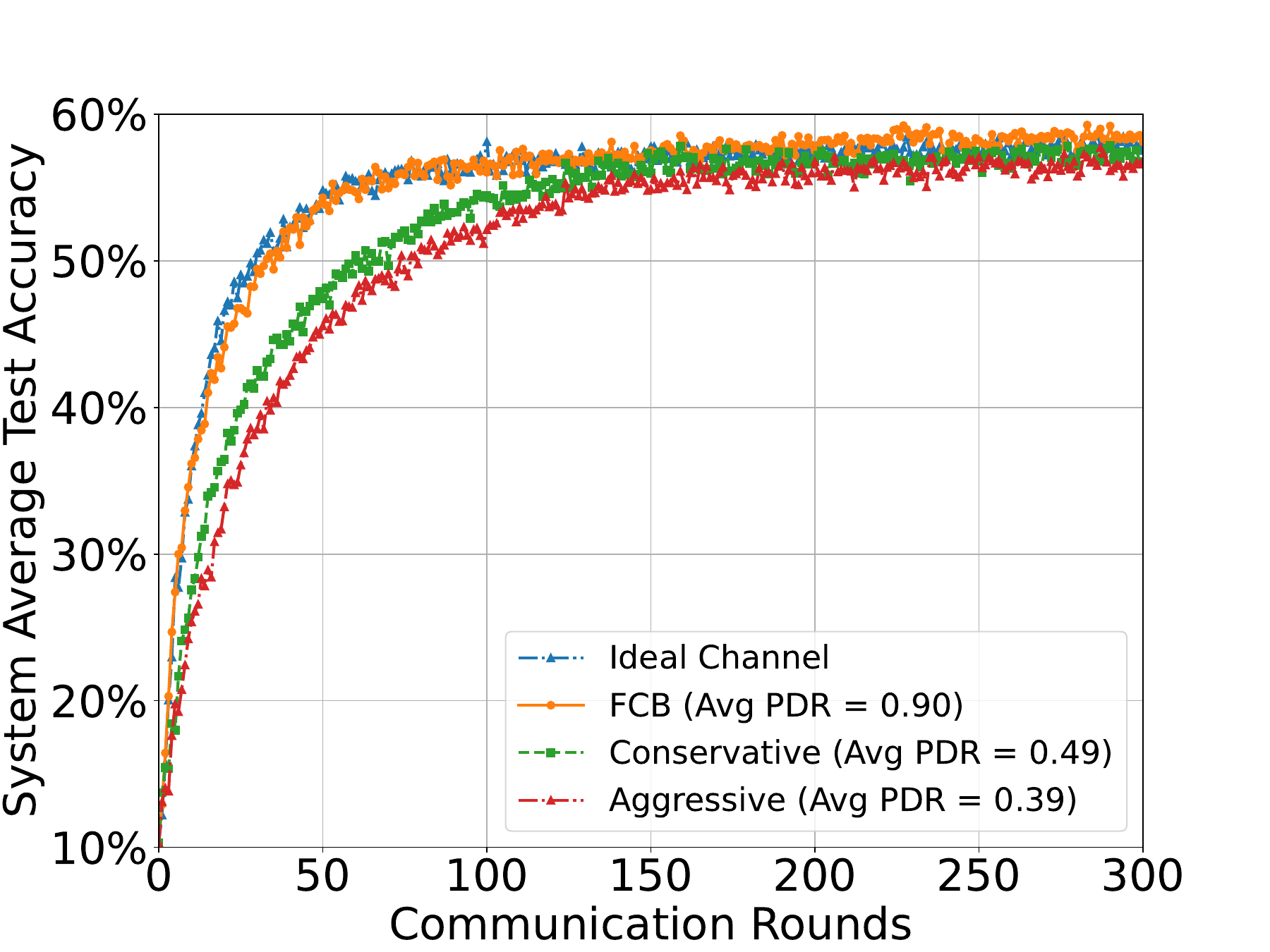}}
    \label{fig:iidsys}
  \end{minipage}
 \begin{minipage}{0.325\textwidth} 
\subfloat[Test accuracy for CIFR-10 mid non-IID case.]{\includegraphics[width=\textwidth]{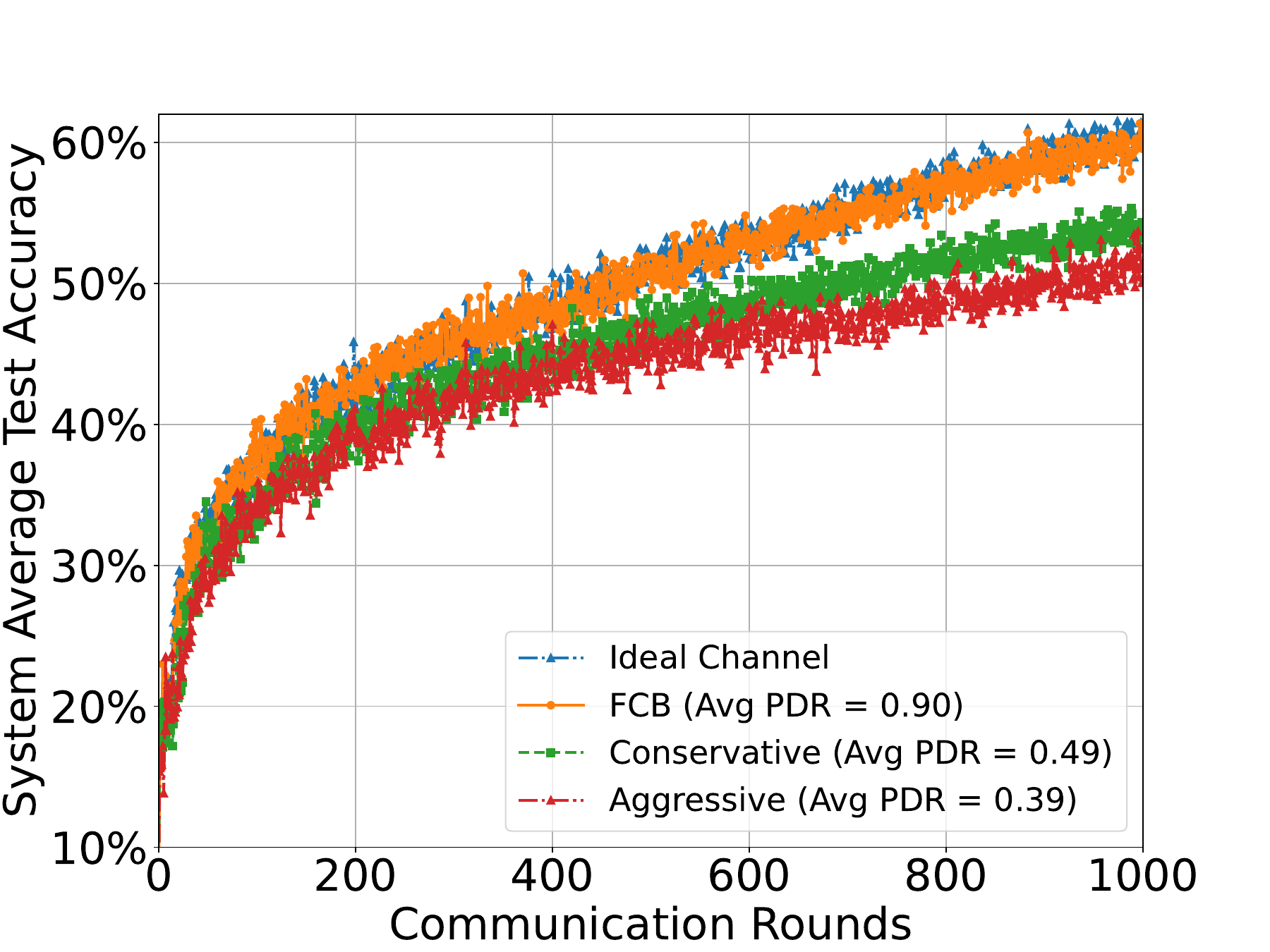}}
    \label{fig:midnonidsys}
  \end{minipage}
 \begin{minipage}{0.325\textwidth}
    \subfloat[Test accuracy for CIFR-10 strong non-IID case.]{\includegraphics[width=\linewidth]{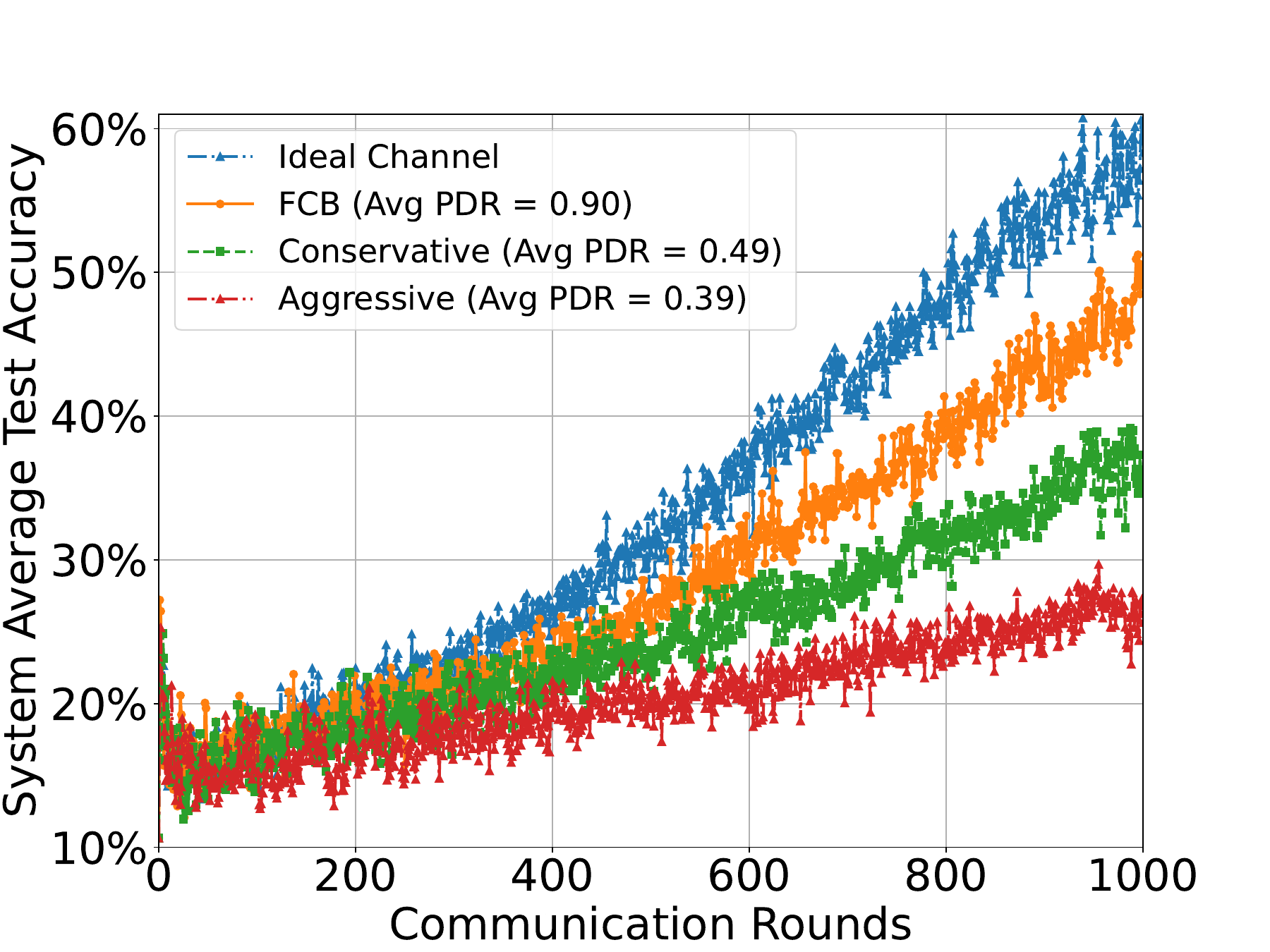}}
\label{fig:noniidsys}
  \end{minipage}
  \caption{System FL performance of CIFAR-10 different data distributions.}
  \label{fig:syst}
\end{figure*}

\noindent
\textbf{FCB Algorithm Results.} Table \ref{tab:fcb_performace} demonstrates that the proposed FCB optimizer outperforms two transmission policy baselines: \textit{1) Aggressive policy} \cite{Guan-2016-ToTransmit}, in which each learner $n$ sets a low threshold $\beta_n$ and therefore prefers to transmit more packets with maximum power $P_n=20$dBm rather than enqueuing them; and \textit{2) Conservative policy} \cite{Channel_Ghazikor_2024}, in which each learner sets $\beta_n$ close to its upper bound $\beta_n^{\max}$ and therefore prefers to enqueue more packets with minimum power $P_n=10$dBm rather than transmitting them. For both $\psi=0.1$ and $\psi=0.4$, FCB achieves the highest average PDR of 0.90 and the highest Jain’s fairness index of 0.99, while maintaining consistently high per-learner PDR values. In contrast, the aggressive baseline suffers from severe interference, resulting in a low average PDR of 0.39 and fairness of 0.80, whereas the conservative baseline improves stability but still achieves only an average PDR of 0.49 and fairness of 0.81. The FCB solution adaptively selects moderate thresholds and powers, avoiding both overly aggressive interference and overly conservative under-utilization of the channel. In addition, although a larger partial observability ($\psi=0.4$) increases the total number of inner-loop CTC iterations from $j=10$ to $j=62$, it converges to the same final solution as $\psi=0.1$, demonstrating the robustness of the FCB algorithm under intermittent participation. Overall, Table \ref{tab:fcb_performace} confirms that FCB achieves better reliability and fairness while remaining robust to partial observability.

\noindent
\textbf{FL Training Results.} We first evaluate the performance under an IID scenario. As shown in Fig. \ref{fig:syst} (a), all schemes converge to a similar final accuracy of about $59\%$, which is expected since each learner has data drawn from the same distribution, leading to consistent local gradients and an accurate approximation of the centralized objective. However, clear differences arise in convergence speed. The FCB algorithm closely follows the ideal-channel benchmark and converges significantly faster than the conservative and aggressive schemes. For instance, to reach $55\%$ accuracy, FCB requires about $45$ communication rounds, while the conservative and aggressive methods require around $65$ and $80$ rounds, corresponding to approximately $30\%$ and $44\%$ faster convergence. This gain is due to improved packet delivery reliability, which reduces information loss and enables more effective aggregation of model updates.

We next consider the mid non-IID scenario, as shown in Fig. \ref{fig:syst} (b). Compared with the IID case, convergence becomes slower and the performance gap between different schemes begins to widen due to moderate data heterogeneity. The ideal-channel benchmark reaches around $60\%$ accuracy, while the FCB algorithm achieves approximately $58\%$, followed by the conservative and aggressive schemes with slightly lower performance. This indicates that even moderate heterogeneity amplifies the impact of packet loss, and reliable transmission becomes increasingly important. The proposed method maintains a consistent advantage, demonstrating its robustness under partially heterogeneous data distributions.

Furthermore, we evaluate the strong non-IID scenario in Fig. \ref{fig:syst} (c). In this case, convergence becomes significantly slower and the final accuracy is further reduced for all schemes due to severe statistical heterogeneity. The ideal-channel benchmark gradually reaches about $60\%$ accuracy, while the FCB algorithm achieves around $50\%$ accuracy, compared to approximately $38\%$ and $28\%$ for the conservative and aggressive schemes, respectively. This corresponds to about $32\%$ and $79\%$ higher final accuracy for FCB. The larger performance gap highlights that packet loss has a more pronounced impact when local data distributions differ significantly. Nevertheless, the FCB consistently outperforms the baselines, demonstrating that the PDR-aware design effectively improves transmission reliability and mitigates the adverse effects of heterogeneity.
\section{Conclusion} \label{conclusion}
In this paper, we investigated UAV-enabled FL in ISM bands, where uplink update transmissions are interference-coupled. Unlike idealized wireless-FL models that assume error-free reception of updates, we considered packetized update delivery, in which buffer overflow, delay violations, and transmission errors can result in partial updates. To capture these effects, we developed a packet-level update delivery model and incorporated the resulting PDR into FL aggregation to reflect partial update reception. Then, we formulated a fairness-consensus bilevel optimization and developed an alternating FCB optimizer composed of the CTC and FPC sub-algorithms, which update threshold and power vectors, respectively, to improve consensus and fairness under partial observability. Numerical results showed that the FCB optimizer enhances the PDR, leading to improved FL aggregation and training performance compared to baseline transmission policies. As future work, we will extend our framework to real-world deployment on software-defined radio (SDR) testbeds, enabling OTA validation of packet-level update delivery over unlicensed ISM bands.

\bibliographystyle{IEEEtran}
\bibliography{ref}

\end{document}